# Advances in the Development of Micropattern Gaseous Detectors with Resistive Electrodes

P. Fonte[1], P. Martinengo[2], E. Nappi[3], R. Oliveira[1], V. Peskov[1,3], F. Pietropaolo[4],

P. Picchi[5]

[1]ISEC and LIP , Portugal
[2]CERN, Geneva , Switerland
[3]INFN Bari, Bari, Italy
[4]Inst. de Ciencias Nucleares  UNAM, Mexico
[5]INFN Padova, Padova, Italy
[6]INFN Frascati, Frascati, Italy

**Abstract**

We describe the most recent efforts made by various groups in implementing resistive electrodes in micropattern gaseous detectors with the aim to combine in the same design the best features of RPCs (for the example, their robustness and spark protection property) with the high granularity and thus the good position resolution offered by microelectronic technology. In the stream of this activity, we have recently developed two novel detectors with resistive electrodes: one was based on resistive micromeshes and the second one is a MSGC with resistive electrodes. We have demonstrated that the resistive meshes are a convenient construction element for various designs of spark-protective detectors: RPCs-type, GEM-type and MICROMEGAS-type. These new detectors enable to considerably enhance the RPC and micropattern detectors applications since they feature not only a high position resolution but also a relatively good energy resolution (25-30 % FWHM at 6 keV) and, if necessary, they can operate in cascaded mode allowing the achievement of a high overall gas gain. The main conclusion from these studies is that the implementation of resistive electrodes in micropattern detectors makes them fully spark protected; on this basis we consider this direction very promising.

1. Introduction

The two major breakthroughs in the field of gaseous detectors after the long and fruitful "wire detectors era" (Geiger counters, proportional counters, MWPCs) are: the development of RPCs [1] and the invention of micropattern detectors (see a review paper [2]). The great success of the RPCs is based on their simplicity, robustness, spark-protection properties and excellent timing characteristics. The exciting improvements made in RPC's performance have been highlighted in several previous RPC conferences.

The main attractive feature of micropattern detectors is that they are manufactured by a modern microelectronic technology, which offers high granularity-and thus unprecedented high position resolution. However, as a consequence of their fine electrode structure, these detectors are very fragile and, for example, could be easily damaged by sparks which may occur during their operation.

By answering the question: "what is the physics behind the spark production and is it possible to avoid them?" it is quiet obvious that in poor and medium quality detectors (which is often the case in practice), due to the small distance between the anode and the cathode structures, various imperfections (sharp edges, dirts and so on) can trigger breakdowns. This is why the requirements on the quality of micropattern detectors are much more demanding than for traditional gases detectors having much thicker avalanche gaps. It is now well established that in high quality detectors operating at low counting rates ($10Hz/mm^2$) the maximum achievable gain is determined by the Raether limit: at $A_m \geq Q_{max}/n_0$ ($A_m$ is the maximum achievable gain, $n_0$ is the number of primary electrons created by ionizing radiation in the detector volume, $Q_{max}$ is a total critical charge in the avalanche, typically $10^6$-$10^7$ electrons) avalanches transform into streamers [3]. At high counting rate, due to some physical effects (see [3] for details) the $A_m$ starts rapidly dropping as the rate increases. The main conclusion from the systematic studies presented in [3] is that discharges in micropattern gaseous detectors during a long-term operation are practically unavoidable. There exist some standard artifices to minimize the destruction of micropattern detectors by sparks (e.g. electrode segmentation in order to reduce the capacitance contributing to the discharges, use detectors in cascaded mode thus increasing $Q_{max}$ and so on), however the implementation of all these expedients had a limited success so far, especially in presence of highly ionizing particles.

In this paper, we describe the recent efforts from various groups to develop micropattern detectors with resistive electrodes. As will be shown, such detectors combine the advantages of RPCs (spark-protection, robustness) to those of micropattern detectors (high granularity and high position resolution). We will also present two novel designs of micropattern detectors with resistive electrodes, recently developed by us: one is based on resistive micromesh and the other on a MSGC with resistive electrodes.

2. Existing micropattern gaseous detectors with resistive electrodes

*2.1.RETGEM*

The first attempt to implement resistive electrodes into micropattern gaseous detectors was described in [4]. The prototype reported in this paper was a thick GEM (see [5]) with electrodes made of graphite paint (widely used for coating outer surfaces of the RPC electrodes). This detector was named Resistive Electrodes Thick GEM or RETGEM. In the subsequent generation of RETGEMs, the electrodes were made of carbonated resistive Kapton 100XC10E5. The RETGEM is in fact a hybrid detector between a RPC and a hole-type gaseous multipliers. Indeed, if the voltage is applied to its resistive electrodes, due to their non infinite resistivity, they will charge up and start acting as equipotential surfaces thus creating the same field lines focusing effect inside the holes as in the case of metallic electrodes. With good quality RETGEMs have been achieved the same maximum gains as with TGEMs (as determined by the Raether limit). However, in contrast to ordinary TGEMs they are fully spark–protected: due to the electrodes resistivity the spark energy was successfully reduced by 100-1000 times [6]. These results were fully confirmed in several independent works [7-9].

Latest advances in RETGEM (RETGEMs with inner metallic meshes or strips) are described in [10].

*2.2.Resistive MICROMEGAS*

The success in spark protection achieved with RETGEMs triggered interest to this approach from other groups. MIGROMEGAS are the other micropattern gaseous detectors to which the resistive electrode approach was applied. In [11] 10 μm thick amorphous silicon or $Si_3N_4$ layers were used to protect a fragile MICROMEGAS pixelized anode plate directly integrated into the SMOS frontend microelectronics (so called Gossip) against the destruction by sparks.

Nowadays, in the framework of the RD51 collaboration, several groups are trying to find the best resistive material for manufacturing the anodes of MICROMEGAS which allow at the same time to preserve the high position resolution of this detector and to make it spark protected [12].

**3. New designs**

*3.1.Gaseous detectors assembled from resistive meshes*

In the stream of the efforts described above, we have developed detectors made of thin resistive meshes manufactured from the resistive Kapton 100XC10E5 (resistivity 2.8-3 MΩ/□, thickness 20μm, hole's diameter 50 μm and hole spacing a 100 μm) manufactured by a laser drilling technique (see Fig. 1). The resistive meshes were stretched either on 5x5 cm$^2$ or 10x10 cm$^2$ G-10 frames. From these stretched meshes, various detectors could be assembled: resistive mesh PPACs (RM-PPAC) with an avalanche gap G=1-3 mm (Fig.2a), resistive mesh MICROMEGAS (RM-μM), G=0.1-0.3 mm (Fig. 2b), resistive mesh GEMs (RM-GEM), G=0.05-0.1mm (Fig. 2c) or several cascaded RM-GEMs or RM-GEM combined with RM- μM. As anodes, we used either metallic plates or G-10 plates covered with resistive Kapton, or (only for position resolution measurements) a ceramic plate with a 50 pitch metallic strips. As a spacer between the meshes and the anode plate or between two meshes, we used a plastic honeycomb structure (for RM-PPAC) or 50-300 μm thick fishing lines, or 50 μm thick meshes made of usual Kapton (for RM-GEMs and RM- μM).
In Fig. 3 are shown the gain vs. voltage curves measured for RM-PPACs and for RM-μM in Ne- and Ar-based mixtures. The measurements were stopped at voltages when first breakdown appear. Due to the small diameter of their holes and the fine pitch, a better position resolution can be achieved with the resistive mesh detectors than with the RETGEMs. The position resolution of this detector was about 300 μm and its energy resolution about 30% FWHM at 6 keV).

The gas gain achieved with RM-GEMs (G=0.05-0.01mm) made from two parallel resistive meshes separated by Kapton spacers were not high (see [13]), most probably due to the spacers and mesh defects. However, RM-μM combined with RM-GEM based preamplifier exhibited an overall gain close to $10^4$ (Fig. 4). No discharge propagation was observed in our experiment when detectors operated in cascade mode. This is because the energy of discharges due to the resistive electrodes was strongly reduced (at least 100 times) and their UV emission becomes too weak to trigger their propagation (see [3] for details).

One of the advantages of adopting the cascade mode is the possibility to reduce an ion back flow to the cathode which can be an attractive feature for some applications such as photodetectors or TPC.

Measurements performed with RM-PPAC at counting rate as high as 5kHz/cm2 did not reveal any changes in pulse amplitudes [13]. As follows from [14] the electrode resistivity can be optimized for higher rates and this will be the subject of the future studies.

*3.2. First prototype of MSGCs with resistive electrodes*

As it looks now, any type of micropattern gaseous detector can be made spark protected by implementing in their design resistive electrodes. To demonstrate the validity of this approach, we recently manufactured and tested first MSGCs prototype with restive electrodes (we call it Resistive MSGC or R-MSGC) produced by the screen printing technique. The width of the anode and the cathode strips were 100 μm and 300μm respectively, their pitch 600 μm. The maximum gain achieved with this prototype (see Fig.5) is currently restricted by the quality of its electrodes (it is much below one expected from the Raether limit) and better quality detectors are under the production now. However, it has already been demonstrated that this R-MSGC is fully spark protected. To achieve higher gains with this particular prototype we used gas preamplification: two parallel resistive meshes with a gap between them 1 mm placed 4 mm above the R- MSGC. In this case overall gains around 1000 were easily achieved (Fig.5). No discharge propagation was observed in our experiments in cascade mode even in pure noble gases which can be an attractive feature in some applications, for example in cryogenic detectors.

## 4. Conclusions

In this paper we summarized the latest developments in various innovative micropattern gaseous detectors with resistive electrodes: RM-PPAC, RM-GEM, RM–MICROMEGAS and R-MSGCs. The main conclusion from their studies is that resistive electrodes make these detectors robust and spark protected opening a new promising direction.

**References:**

[1] R. Santonico, Nucl. Phys B (Proc. Suppl), 158, 2006, 5 and Nucl. Instr. and Meth, A602, 2009, 627
[2] V. Peskov, arXiv:0906.5215, 2009,IEEE pproceedings, DO I: 10.1109/IWASI.2009.5184767
[3] V. Peskov et al., arXiv:0911.0463, 2009
[4] J.M. Bidault et al., Nucl. Phys B (Proc. Suppl), 158, 2006, 199
[5] L. Periale et al., Nucl. Instr. and Meth., A478, 2002, 377
[6] A. Di Mauro et al., arXiv:0706.0102, 2007

[7] R. Akimoto et al , Report at the Micropattern Conference MPGD2009, Crete, Greece, http://indico.cern.ch/materialDisplay.py?contribId=71&sessionId=6&materialId=slides&confId=50396.
[8] A. Bondar et al., *JINST 3 P07001*, 2008
[9] V. Razin et al., arXiv:0911.4807, 2009
[10] P. Fonte et al., Nucl. Instr. and Meth, A602, 2009, 850
[11] H. Van der Graaf, Nucl. Instr. and Meth,, A607, 2009, 78
[12] See reports at the RD51 Collaboration meeting, CERN, Dec. 2009, http://indico.cern.ch/conferenceDisplay.py?confId=72610
[13] R. Oliveira et al., arXiv:1002.1415, 2010
[14] P. Fonte etal., Nucl. Instr. and Meth, A431,1999, 154

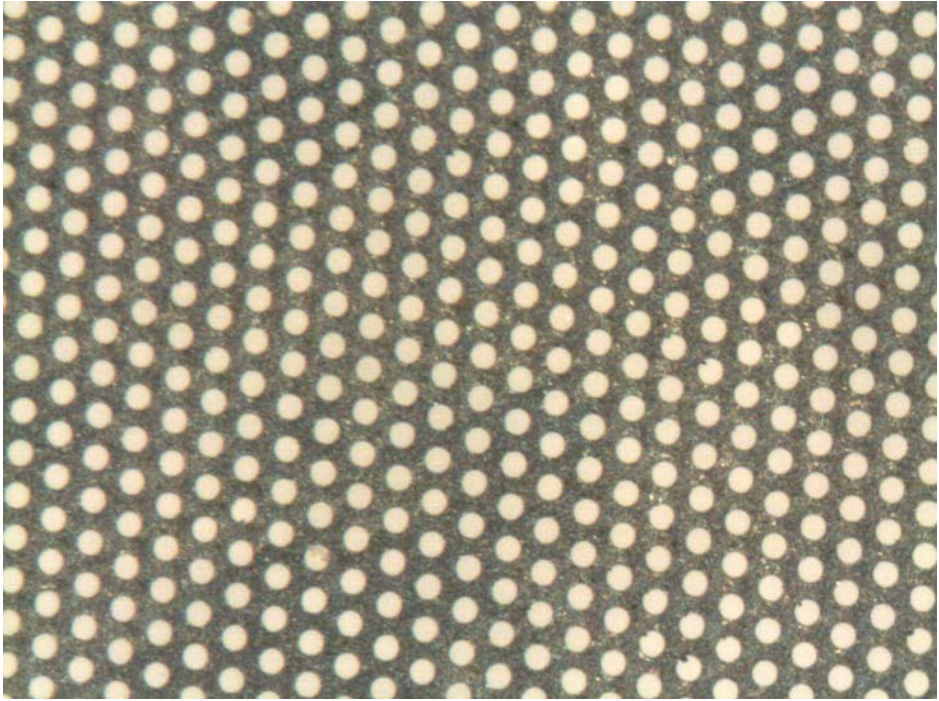

Fig.1. A magnified photograph of restive mesh used in this work for construction various gaseous detectors

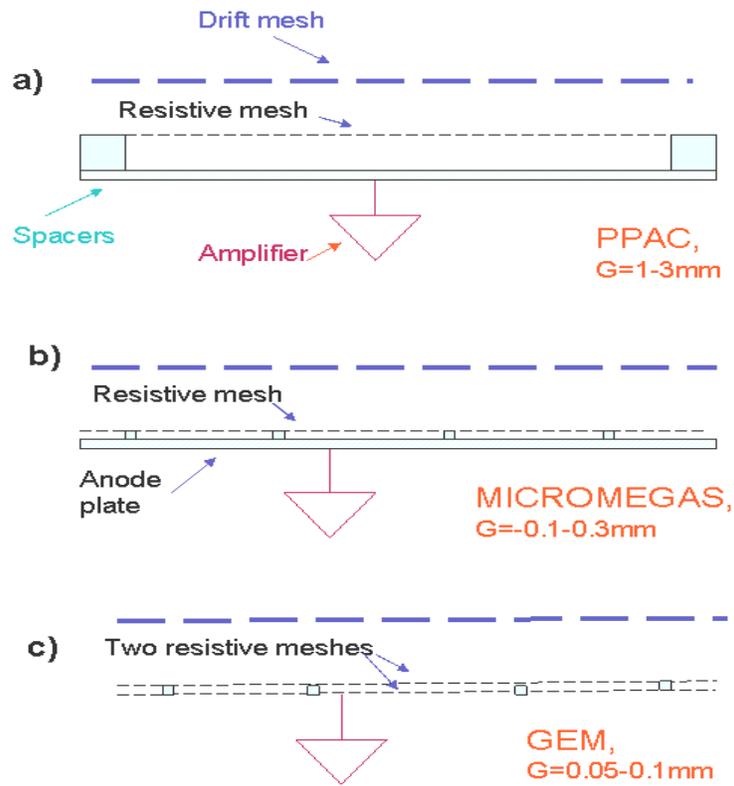

Fig.2. Schematic drawing of detectors assembled from resistive meshes: a )RM-PPAC, b)RM-MICROMEGAS, c) RM-GEM,

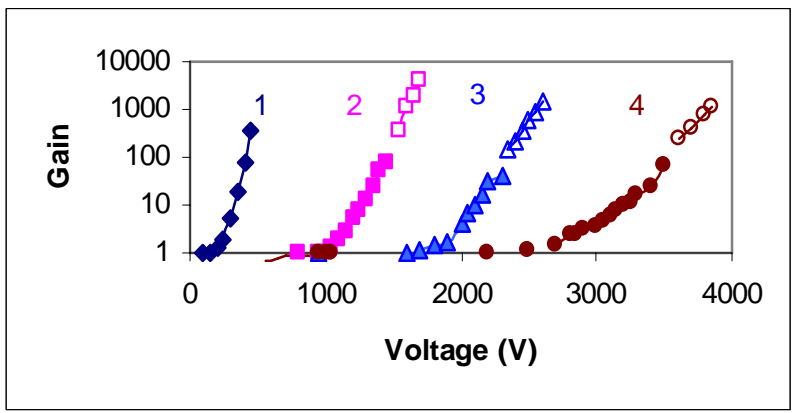

Fig.3 Gain curves for RM-MICROMEGAS and RM-PPACs:1-G=0.15mm, Ar+15%$CO_2$, 2-G=1mm, Ne+8%$CH_4$, 3-G=1mm, Ar+8%$CH_4$, 4-G=3.5mm, Ne+12%$CH_4$. Filled symbols measurements with alphas, open symbols- $^{55}$Fe.

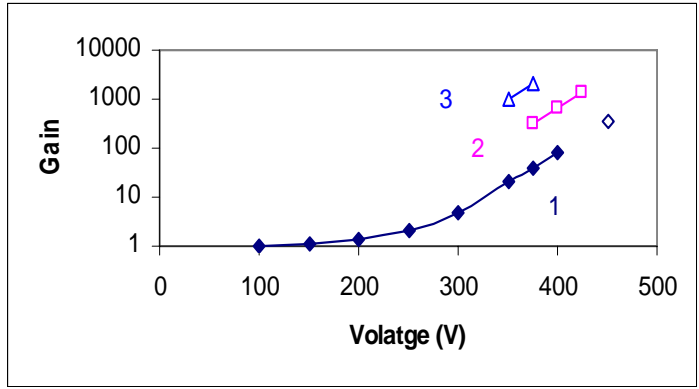

Fig.4 Gain curves for RM-MICROMEGAS, G=1=0.15mm (curve 1) and for RM-MICROMEGAS combined with RM-GEM (G=0.25mm) gas preamplifier operating at 600V across it (2) and 700V(3) in Ar+15%$CO_2$. Filled symbols-alphas, open symbols-$^{55}$Fe.

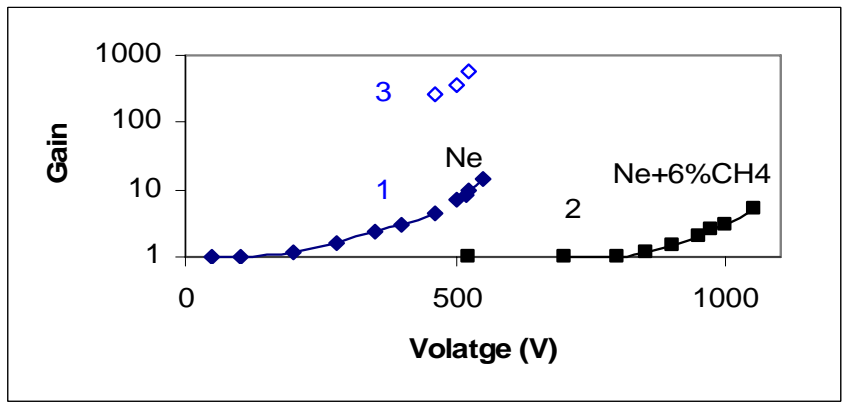

Fig.5. Gain vs. the voltage for R-MSGC operating in Ne (1,3) and Ne+6%$CH_4$(2). Curve 3 corresponds to the overall gain of R-MSGC combined with RM-GEM, G=1mm, operating at applied voltage of 300V.